\documentclass[prl,aps,twocolumn,showpacs, floatfix]{revtex4}
\usepackage{amssymb}
\usepackage{hyperref}
\usepackage{graphicx}
\usepackage{dcolumn}
\usepackage{bm,amsmath,verbatim}

\def\be{\begin{equation}}
\def\ee{\end{equation}}
\def\bea{\begin{eqnarray}}
\def\eea{\end{eqnarray}}
\def\bse{\begin{subequations}}
\def\ese{\end{subequations}}

\def\be{\begin{eqnarray}}
\def\ee{\end{eqnarray}}

\begin{document}

\title{1D topological chains with Majorana fermions in 2D non-topological
optical lattices}
\author{Lei Jiang, Chunlei Qu}
\author{Chuanwei Zhang}
\thanks{Corresponding author. \\
Email: chuanwei.zhang@utdallas.edu}
\affiliation{Department of Physics, The University of Texas at Dallas, Richardson, Texas
75080, USA}

\begin{abstract}
The recent experimental realization of 1D equal Rashba-Dresselhaus
spin-orbit coupling (ERD-SOC) for cold atoms provide a disorder-free
platform for the implementation and observation of Majorana fermions (MFs),
similar to the well studied solid state nanowire/superconductor
heterostructures. However, the corresponding 1D chains of cold atoms possess
strong quantum fluctuation, which may destroy the superfluids and MFs. In
this Letter, we show that such 1D topological chains with MFs may be on
demand generated in a 2D non-topological optical lattice with 1D ERD-SOC by
modifying local potentials on target locations using experimentally already
implemented atomic gas microscopes or patterned (e.g., double or triple
well) optical lattices. All ingredients in our scheme have been
experimentally realized and the combination of them may pave the way for the
experimental observation of MFs in a clean system.
\end{abstract}

\date{\today }
\pacs{03.75.Ss, 05.30.Fk, 03.65.Vf, 67.85.-d}
\maketitle

Majorana fermions (MFs) \cite{wilczek09} obey non-Abelian exchange
statistics and are crucial for realizing fault-tolerant topological quantum
computation \cite{nayak08,hasan10,qi11}. Following initial theoretical
proposals \cite{volovik88,read91,kitaev01,read00,Das
Sarma06,fu08,sau10,alicea10,sau10b,lutchyn10,oreg10}, some possible
signatures of MFs have been observed recently in experiments \cite%
{mourik12,deng12,das12,Rokhinson12,Veldhorst12,finck13,yazdani14} using 1D
nanowires or ferromagnetic atomic chains on top of an \textit{s}-wave
superconductor and with strong spin-orbit coupling (SOC). However, these
signatures are not conclusive because of disorder and other complications in
solid state \cite{kells12,lee12,Das Sarma12,lieber12,beenakker12,churchill13}%
. In this context, the recent experimental realization of SOC \cite%
{lin11,pan12,engles13,chen14,zhang12,Zwierlein12} in ultra-cold atomic gases
provides a disorder-free and highly controllable platform for observing MFs.
In experiments, 1D equal Rashba-Dresselhaus SOC (ERD-SOC) and tunable Zeeman
field have been achieved, which, together with \textit{s}-wave
superfluidity, make it possible to observe MFs \cite%
{sato09,zhu11,seo12,gong12,liu12,jiang11,mueller12,liu13} in 1D atomic tubes
or chains, similar as the nanowire systems.

However, unlike solid state nanowire systems where \textit{s}-wave
superconducting pairs are induced from proximity effects, the superfluid
pairing in the 1D atomic chain is self-generated from the \textit{s}-wave
contact interaction, leading to the strong quantum fluctuation that renders
the long range superfluid order impossible in the thermodynamic limit. To
circumvent this obstacle, quasi-1D systems with multiple weakly coupled
uniform chains \cite%
{wimmer10,tewari12b,qu13,li13,wang14,sato13,oreg14,oreg14b,Nagaosa14} have
been studied in both solid state and cold atoms, where transverse tunneling
was found to lift the zero energy degeneracy of multiple MFs.

In this Letter, we consider a truly 2D non-topological fermionic optical
lattice with the experimentally realized 1D ERD-SOC. We raise the question
whether single or multiple topological 1D chains supporting MFs can be on
demand generated at target locations in such non-topological 2D systems.
Generally, a 1D chain in a 2D lattice can be locally modified to satisfy the
topological condition for MFs using the recently experimentally realized
single site addressing (the atomic gas microscopes) \cite%
{wurtz09,greiner09,greiner10,bloch10,bloch11} or patterned (e.g., double or
triple well) optical lattices \cite{porto06,folling07,Ketterle13}. However,
the atom chain is strongly coupled with neighboring chains through
transverse tunneling in the 2D system, therefore a naive expectation is that
the coupling may destroy the local topological properties and the associated
MFs.

Here we show that 1D topological chains with MFs can indeed be generated on
demand from 2D non-topological fermionic optical lattices with
experimentally achieved 1D ERD-SOC. Local addressing lasers in atomic gas
microscopes can modify the effective local chemical potentials along single
or multiple 1D chains, leading to a quantum transition to discrete
topological chains that are characterized by non-zero winding numbers and
host MFs at chain ends. Multiple MFs in spatially separated multiple
topological chains still couple, with the coupling induced zero energy
splitting exponentially decaying with the distance of neighboring
topological chains. Note that similar results apply also to 3D if 1D chains
can be locally addressed in a 3D optical lattice. In the weak transverse
tunneling region (quasi-1D), the MF coupling is extremely small for two
topological chains separated by one or two non-topological chains, making it
possible to observe multiple MFs in 2D or 3D double or triple well optical
lattices without requiring the single site addressing.

\begin{figure}[t]
\includegraphics[scale=0.3]{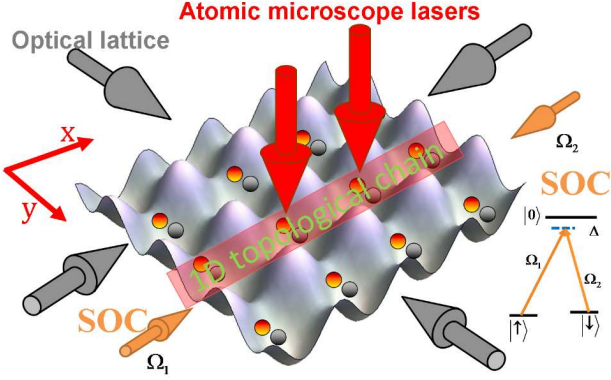}
\caption{Illustration of the proposed experimental setup. Grey arrows
represents 2D square optical lattice lasers. Red tube demonstrates the 1D
potential chain induced by local addressing lasers in atomic microscopes
(red arrows). Two counter-propagating Raman lasers (orange arrows) couple
two spin states, generating 1D ERD-SOC \protect\cite%
{lin11,pan12,engles13,chen14,zhang12,Zwierlein12}. }
\label{f0}
\end{figure}

\emph{Model:} We consider a spin-$1/2$ ultra-cold degenerate fermionic gas
(spin $\uparrow $ and $\downarrow $) in a 2D square lattice with the lattice
size $N=n_{x}\times n_{y}$. As shown in the schematic picture Fig. \ref{f0},
two Raman lasers couple two spin states to induce 1D ERD-SOC along the
\textit{x}-axis. The far-detuned local addressing lasers \cite%
{wurtz09,greiner09,greiner10,bloch10,bloch11} can modify the local potential
of the optical lattice along a 1D chain at target locations along the
\textit{x}-direction. Multiple spatially well separated 1D chains can also
be generated using additional local addressing lasers. In the 2D lattice,
the tight-binding mean-field BdG Hamiltonian
\begin{equation}
H_{BdG}=H_{L}+H_{SOC}+H_{D}+H_{\Delta },  \label{BDGH}
\end{equation}%
where $H_{L}=-\sum\nolimits_{\mathbf{i},\sigma ,\eta }t_{\eta }\left( C_{%
\mathbf{i},\sigma }^{\dagger }C_{\mathbf{i}+\mathbf{\eta },\sigma
}+h.c.\right) -\sum\nolimits_{\mathbf{i},\sigma }\bar{\mu}C_{\mathbf{i}%
,\sigma }^{\dagger }C_{\mathbf{i},\sigma }$ is the bare Hamiltonian in the
2D lattice with $\eta =\left\{ x,y\right\} $. The fermionic operator $C_{%
\mathbf{i},\sigma }^{\dagger }$ ($C_{\mathbf{i},\sigma }$) creates
(annihilates) a particle with spin $\sigma $ at site $\mathbf{i}%
=(i_{x},\,i_{y})$. We use $\bar{\mu}=\mu -2t_{x}-2t_{y}$ for the effective
chemical potential to match with that in the continuous model.\textbf{\ }$%
t_{x}$ and $t_{y}$ are the nearest neighbor tunnelings along \textit{x} and
transverse \textit{y} directions respectively\textbf{.} $H_{SOC}=\alpha
\sum\nolimits_{\mathbf{i}}(C_{\mathbf{i},\uparrow }^{\dagger }C_{\mathbf{i}-%
\hat{x},\downarrow }-C_{\mathbf{i},\uparrow }^{\dagger }C_{\mathbf{i}+\hat{x}%
,\downarrow }+H.c.)+h_{z}\sum_{\mathbf{i}}(C_{\mathbf{i},\uparrow }^{\dagger
}C_{\mathbf{i},\uparrow }-C_{\mathbf{i},\downarrow }^{\dagger }C_{\mathbf{i}%
,\downarrow })$ describes the experimentally available 1D ERD-SOC, with the
SOC strength $\alpha $ and Zeeman field $h_{z}$. $H_{D}=\sum_{\mathbf{i}%
,\sigma }V_{T}(i_{y})C_{\mathbf{i},\sigma }^{\dagger }C_{\mathbf{i},\sigma }$
represents the 1D potential dip with the local potential $V_{T}(i_{y})$,
which is generated by local addressing lasers and non-zero only at the $i_{y}
$ chains. $H_{\Delta }=-\sum\nolimits_{\mathbf{i}}\Delta _{\mathbf{i}}(C_{%
\mathbf{i},\uparrow }^{\dagger }C_{\mathbf{i},\downarrow }^{\dagger }+h.c.)$
is the mean-field paring Hamiltonian, with the order parameter $\Delta _{%
\mathbf{i}}=-g\langle C_{\mathbf{i},\downarrow }C_{\mathbf{i},\uparrow
}\rangle $ and on-site interaction strength $g$. Hereafter we take $t_{x}=t$
as the energy unit. We solve the corresponding BdG equation
self-consistently with the pairing gap equation and fixed chemical
potential, following the standard numerical procedure \cite%
{qu13b,xu14,qu14,xu14b,jiang14}. The order parameter is chosen to be complex
to find the ground state. We use the box boundary condition for the
self-consistent calculation. In practical experiment, there is a weak
harmonic confinement which may change the location of MFs, but don't change
the essential physics \cite{jiang14}.

The above BdG Hamiltonian preserves particle-hole symmetry $\Xi H_{BdG}\Xi
^{-1}=-H_{BdG}$, where $\Xi =\tau _{x}\mathcal{K}$, $\tau _{x}=\tilde{\tau
_{x}}\otimes \tilde{\sigma _{0}}\otimes \tilde{\rho _{0}}$, $\tilde{\tau _{i}%
}$, $\tilde{\sigma _{i}}$ are $2\times 2$ Pauli matrices acting on
particle-hole and spin spaces respectively, $\tilde{\rho _{0}}$ is a $%
N\times N$ identity matrix on the lattice site space, and $\mathcal{K}$ is
the complex conjugate operator. If the order parameter $\Delta _{i}$ is
real, the Hamiltonian is also real, which preserves a time-reversal like
symmetry $\Theta H_{BdG}\Theta ^{-1}=H_{BdG}$ with $\Theta =\mathcal{K}$, as
well as a chiral symmetry $\mathcal{S}H_{BdG}\mathcal{S}^{-1}=-H_{BdG}$ with
$S=\Theta \cdot \Xi =\tau _{x}$. In this case, the system belongs to the BDI
topological class characterized by a $\mathbb{Z}$ topological invariant \cite%
{schnyder08,teo10}.

\emph{One topological chain:} First consider a 2D lattice with no tunneling
along the \textit{y}-axis ($t_{y}=0$), thus the 2D lattice is composed of
individual $x$-direction 1D chains. At the central chain we add an extra
constant potential $V_{_{T}}(y_{c})=V$, so that the central chain becomes
topological, while other parts of the system are still in the
non-topological region. Here the topological region is defined locally by
the criteria $h_{z}\geqslant \sqrt{(\mu -V_{T})^{2}+\triangle ^{2}}$, same
as the usual 1D topological chains \cite{sau10,oreg10}. In our numerical
calculations, we take $n_{x}=81$, $n_{y}=9$, and $y_{c}=5$. In the central
topological chain, two MFs should exist at two ends. When $t_{y}$ is
increased from zero to $t$, the topological chain couples with neighboring
non-topological chains and the system changes from 1D to quasi-1D and
finally to 2D. A nature question is whether MFs at the center chain will
survive with the strong coupling.

\begin{figure}[tbp]
\includegraphics[scale=0.8]{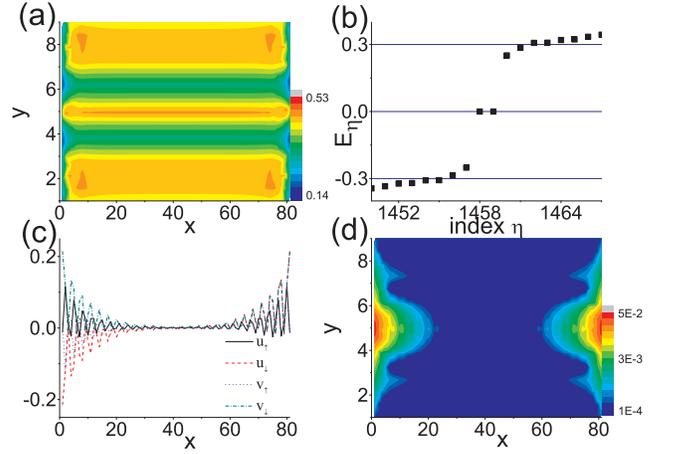}
\caption{MFs in single 1D topological chain in the 2D optical lattice. (a)
The amplitude of the order parameter $|\Delta |$. (b) The quasiparticle
energy spectrum. (c) The zero energy mode wave function along the central
chain. (d) Zero energy mode density (Log scale). Parameters: $t_{y}=t
$, $\protect\alpha =2t$, $g=-5.5t$, $h_{z}=1.4t$, $V=-1.45t$, $\protect\mu %
=-1.555t$.}
\label{f1}
\end{figure}

Fig. \ref{f1} demonstrates the existence of MFs even in truly 2D region with
$t_{y}=t$. The amplitude of the superfluid order parameter $\Delta _{\mathbf{%
i}}$ is plotted in Fig. \ref{f1}(a). We find that $\Delta _{\mathbf{i}}$ is
homogeneous along the \textit{x}-axis in the self-consistent calculation,
except at the boundary. Furthermore, $\Delta _{\mathbf{i}}$ has a constant
phase across on the whole 2D system, therefore we can choose it to be real
without loss of generality for the discussion of the topological properties.
Fig. \ref{f1}(b) shows the quasiparticle energy spectrum, where we clearly
see the existence of Majorana zero energy modes with a tiny energy splitting
$E\approx \pm 2\times 10^{-5}t$ mainly due to the finite size effect. The
mini-gap energy, defined as the energy difference between the zero energy
mode and the next lowest quasiparticle state, is comparable to the order
parameter $\Delta _{\mathbf{i}}$. Fig. \ref{f1}(c) shows the wavefunction of
the zero energy mode ($E\approx +2\times 10^{-5}t$ state) along the central
chain, which satisfies the relation for MFs: $u_{\sigma }=\lambda v_{\sigma }
$,$\,\lambda =\pm 1$, indicating the central chain is still topological with
two MFs at its ends. Fig. \ref{f1} (d) shows the density of the zero energy
mode, which is square of the zero energy mode wave function for both spin up
and spin down atoms in the 2D plane. The zero energy mode still localizes at
the ends of the central chain, but slightly spread to neighboring chains
which is due to the transverse tunneling. Note that in practical
experiments, there exists a finite detuning for the Raman coupling between
two bare spin states, which corresponds to an in-plane Zeeman field $%
h_{y}\sigma _{y}$ in our notation. Such non-zero in-plane Zeeman field is
known to break the inversion symmetry and lead to the FF type of ground
states with finite momentum pairing. We confirm that our results still hold
in the FF state (see Supplemental Material).

\emph{Topological characterization}: The emergence of MFs at the edges of
the central chain originates from the bulk topological properties of the 2D
optical lattice with the imprinted 1D topological chain. In the above
self-consistent BdG calculations, both order parameter and atom density are
almost homogeneous along the \textit{x}-axis, therefore it would be a good
approximation to assume that the bulk is uniform. With a periodic boundary
condition along the $x$-axis, the momentum $k_{x}$ is a good quantum number.
The 2D lattices can be taken as a series of individual 1D chains coupled
through transverse tunneling, with the effective BdG Hamiltonian%
\begin{equation}
H_{BdG}(k_{x})=H_{0}(k_{x})\rho _{0}+(V\tau _{z}\sigma _{0}+\triangle
^{\prime }\tau _{y}\sigma _{y})\rho ^{\prime }-t_{y}\tau _{z}\sigma _{0}\rho
_{x},  \label{BDG2}
\end{equation}%
where $H_{0}(k_{x})=[-2t_{x}\cos k_{x}-\bar{\mu}]\tau _{z}\sigma
_{0}+2\alpha \sin k_{x}\tau _{z}\sigma _{y}+h_{z}\tau _{z}\sigma _{z}+\Delta
_{0}\tau _{y}\sigma _{y}$ describes the original uniform individual chains
with the SOC and Zeeman field. $\rho $ spans the \textit{y}-axis chain space
with $\rho _{0}$ as the identity matrix. $(\rho ^{\prime })_{ij}=1$ for $%
i=j=y_{c}$ and $0$ otherwise. The $\rho ^{\prime }$ part describes the
potential and order parameter differences of the central chain from others.
The $\rho _{x}$ term describes the \textit{y}-axis hopping between nearest
neighboring chains, with $(\rho _{x})_{i,j}=1$ for $|i-j|=1$ and $0$ for
others ($i,j=1\cdots n_{y}$).

The topological properties of the BdG Hamiltonian (\ref{BDG2}) can be
characterized by the winding number $W$. For a single 1D chain in the 2D
optical lattice, the BdG Hamiltonian is in the BDI topological class with a
chiral symmetry $SH_{BdG}(k_{x})S^{-1}=-H_{BdG}(k_{x})$, where $S=\tau _{x}$%
. Therefore the BdG Hamiltonian can be transformed to be off-diagonal in the
$\tau $ space
\begin{equation}
UH_{BdG}U^{+}=\left[
\begin{array}{cc}
0 & A(k_{x}) \\
A^{T}(-k_{x}) & 0%
\end{array}%
\right] =h(k_{x})\tau _{x}+d\tau _{y}
\end{equation}%
by a unitary transformation $U=e^{-i(\pi /4)\tau _{y}}$, where $%
A(k_{x})=h(k_{x})-id$, $h(k_{x})=\{[-2t_{x}\cos k_{x}-\bar{\mu}]\sigma
_{0}+2\alpha \sin k_{x}\sigma _{y}+h_{z}\sigma _{z}\}\rho _{0}-t_{y}\sigma
_{0}\rho _{x}+V\sigma _{0}\rho ^{\prime }$, and $d=\Delta _{0}\sigma
_{y}\rho _{0}+\triangle ^{\prime }\sigma _{y}\rho ^{\prime }$. The winding
number
\begin{equation}
W=-\frac{i}{\pi }\int_{k_{x}=0}^{\pi }\frac{dz}{z(k_{x})},  \label{winding}
\end{equation}%
where $z(k_{x})=\det A(k_{x})/|\det A(k_{x})|$ \cite{tewari12}.

\begin{figure}[tbp]
\includegraphics[scale=0.42]{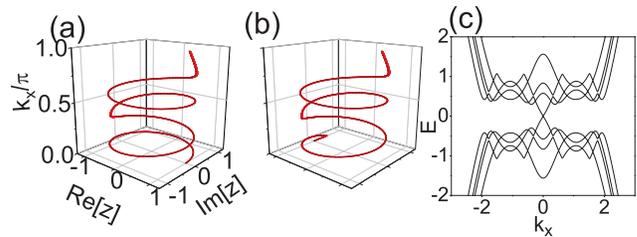}
\caption{Complex value $z$ as function of $k_{x}$ in the region $(0,\protect%
\pi )$ with (a) $V=-0.8t$ (non-topological) and (b) $V=-1.2t$ (topological).
The other parameters: $t_{y}=t$, $\protect\alpha =2t$, $\Delta _{0}=0.4t$, $%
h_{z}=1.4t$, $\triangle ^{\prime }=0.1t$, $\protect\mu =-1.555t$. (c) Band
structure at the topological phase transition point $V=-t$.}
\label{f3}
\end{figure}

Fig. \ref{f3} shows the change of the topological properties with the
potential $V$ along the central chain. When $V$ is small, the whole lattice,
including the center chain, is in the non-topological region. The
corresponding complex value of $z$ rotates when $k_{x}$ changes from $0$ to $%
k_{x}=\pi $ as shown in Fig. \ref{f3} (a), indicating $W=0$. When the
potential depth $|V|$ increases beyond a threshold value $V_{c}\approx -t$, $%
W$ changes to $-1$. Across $V_{c}$, the band gap closes (Fig. \ref{f3} (c))
and reopens, indicating a topological phase transition to a phase where the
central chain becomes topological and hosts a pair of Majorana fermions,
agreeing with the self-consistent calculation.

\begin{figure}[t]
\includegraphics[scale=0.8]{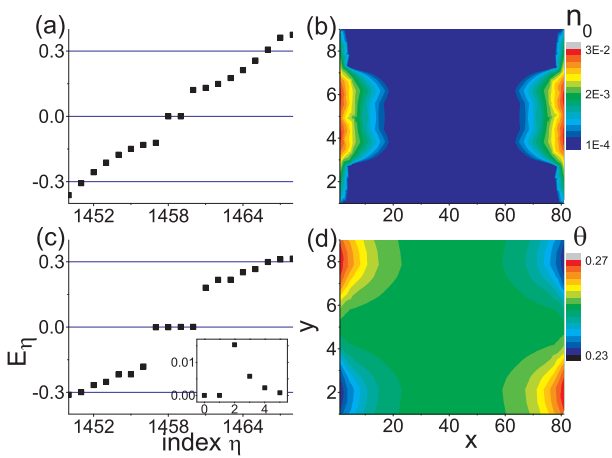}
\caption{MFs in two topological chains at $y=4$ and $y=6$ (a),(b) or at $y=2$
and $y=8$ (c),(d). (a) Quasiparticle energy spectrum for two topological
chains separated by one non-topological chain. (b) Density of the zero
energy mode (Log scale). (c) Quasiparticle energy spectrum for two
topological chains separated by five non-topological chains. The inset plots
the lowest positive quasiparticle energy as a function of the number of
non-topological chains in between. (d) The phase of the order parameter $%
\Delta $. $t_{y}=t$, the other parameters are the same as Fig. \protect\ref%
{f1}. }
\label{f4}
\end{figure}

\emph{Multiple topological chains:} Multiple topological chains may be
generated using additional local addressing lasers to obtain multiple MFs.
We first consider two topological chains separated by one non-topological
chain, with the extra potential $V_{T}$ adding at $y=4$ and $y=6$. The
energy spectrum from the self-consistent BdG calculation is plotted in Fig. %
\ref{f4} (a), showing one zero energy mode although there are two
topological chains. This is due to the strong coupling between two chains,
leading to the winding number $W=-1$, instead of $-2$. Therefore there is
only one pair of MFs when two chains are close. Fig. \ref{f4} (b) shows the
density of the zero energy mode, which widely spreads along the \textit{y}%
-axis from $y=4$ to $y=6$.

When two topological chains are further separated by more than one
non-topological chains, the analysis of the bulk topological properties
based on constant order parameter phase shows that $W=-2$, indicating two
pairs of MFs. However, in the self-consistent calculation, the order
parameter phase is no longer uniform due to the interaction between two MFs
at the same end, leading to the splitting of the zero energy states. In Fig. %
\ref{f4} (c), we plot the quasiparticle energy spectrum for two topological
chains located at $y=2$ and $y=8$ obtained from the self-consistent
calculation. Fig. \ref{f4} (d)) shows the phase $\theta (x,y)$ of order
parameter $\Delta =|\Delta |e^{i\theta (x,y)}$, which has an antisymmetric
structure between two topological chains. The phase difference between two
ends of one topological chain is opposite from the other topological chain.

In principle there should be no zero energy modes left due to the
interaction between MFs, which splits the energy away from zero to a finite
value \cite{li13,wang14}. In practice, due to the large distance between two
topological chains, the coupling strength between two MFs on different
topological chains is extremely small and the energy splitting becomes
negligible. For instance, the five chain separation in Fig. \ref{f4} (c)
leads to an energy splitting $E\approx 8\times 10^{-4}t$. The inset in Fig. %
\ref{f4} (c) shows the change of the splitting with the number of
non-topological chains between two topological ones. When the number is 0
and 1, $W=-1$, and the splitting is almost zero since there is only one MF
at each end. When the number is 2 and above, the winding number becomes $-2$
and the interaction of two MFs induces a splitting. The energy splitting
decreases exponentially with the distance between two topological chains and
approaches almost zero for the 5 lattice separation.

\begin{figure}[tbp]
\includegraphics[scale=0.8]{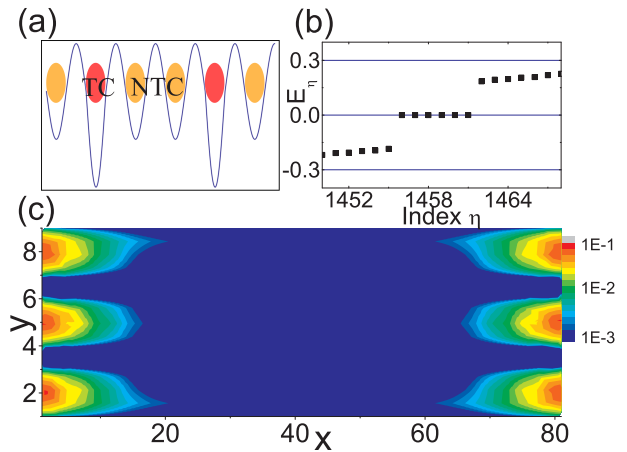}
\caption{MFs in quasi-1D triple-well superlattices. (a) Illustration of the
triple-well superlattices along the $y$-axis. Topological chains (TC) are at
$y=2$, $y=5$ and $y=8$, separated by non-topological chains (NTC). (b)
Quasiparticle energy spectrum showing three pairs of zero energy modes. (c)
Density of zero energy modes (Log scale). Other parameters $%
t_{y}=0.1t$, $\protect\alpha =0.75t$, $g=-3.5t$, $h_{z}=0.7t$, $V=-0.55t$, $%
\protect\mu =-0.55t$.}
\label{f5}
\end{figure}

\emph{Multiple MFs in superlattices:} The interaction between MFs in
topological chains can also be significantly reduced by decreasing the
tunneling along the transverse direction, which makes the system quasi-1D,
instead of 2D. In this case, no large separation between neighboring
topological chains is needed, making it possible to generate multiple MFs
using patterned optical superlattices along the \textit{y}-axis. In
experiments, optical superlattices such as double well or triple well
lattices can be generated using the superposition of different lattice beams
\cite{porto06,folling07,Ketterle13}, which are much easier than the single
site addressing. Fig. \ref{f5} (a) shows a triple well optical lattice along
the \textit{y}-axis with one of the triple wells in the topological region
(i.e., two neighboring topological chains separated by two non-topological
chains). With a small transverse tunneling $t_{y}=0.1t$, the energy
splitting for the MF zero energy state is as tiny as $E\approx 5\times
10^{-5}t$, as shown in Fig. \ref{f5}. In our calculation, we put three
topological chains at $y=2$, $y=5$ and $y=8$, and find one pairs of MFs
formed at each topological chain ends. We also confirm that similar physics
occurs if another triple well superlattice is applied along the $z$-axis to
form a 3D lattice with weak tunneling along both $y$ and $z$ directions.

\emph{Conclusion:} We show that, with the assistant of atomic gas
microscopes or patterned optical super-lattices, 2D non-topological optical
lattices with experimentally achieved 1D ERD-SOC can host non-coupled 1D
topological chains with MFs. We emphasize that although we illustrate the
idea using a 2D geometry, the same physics also applies to 3D
non-topological optical lattices, providing selected 1D chains can be
locally modified or patterned optical superlattices are used. All
ingredients in our proposed schemes are available in current experiments,
and the scheme may lead to unambitious experimental signature of the
long-sought MFs in a clean cold atomic system.

\begin{acknowledgments}
\textbf{Acknowledgements}: This work is supported by ARO (W911NF-12-1-0334) and AFOSR (FA9550-11-1-0313
and FA9550-13-1-0045).
\end{acknowledgments}

\newpage

\section*{Supplemental Material}

\begin{figure}[htbp]
\includegraphics[width=0.45\textwidth]{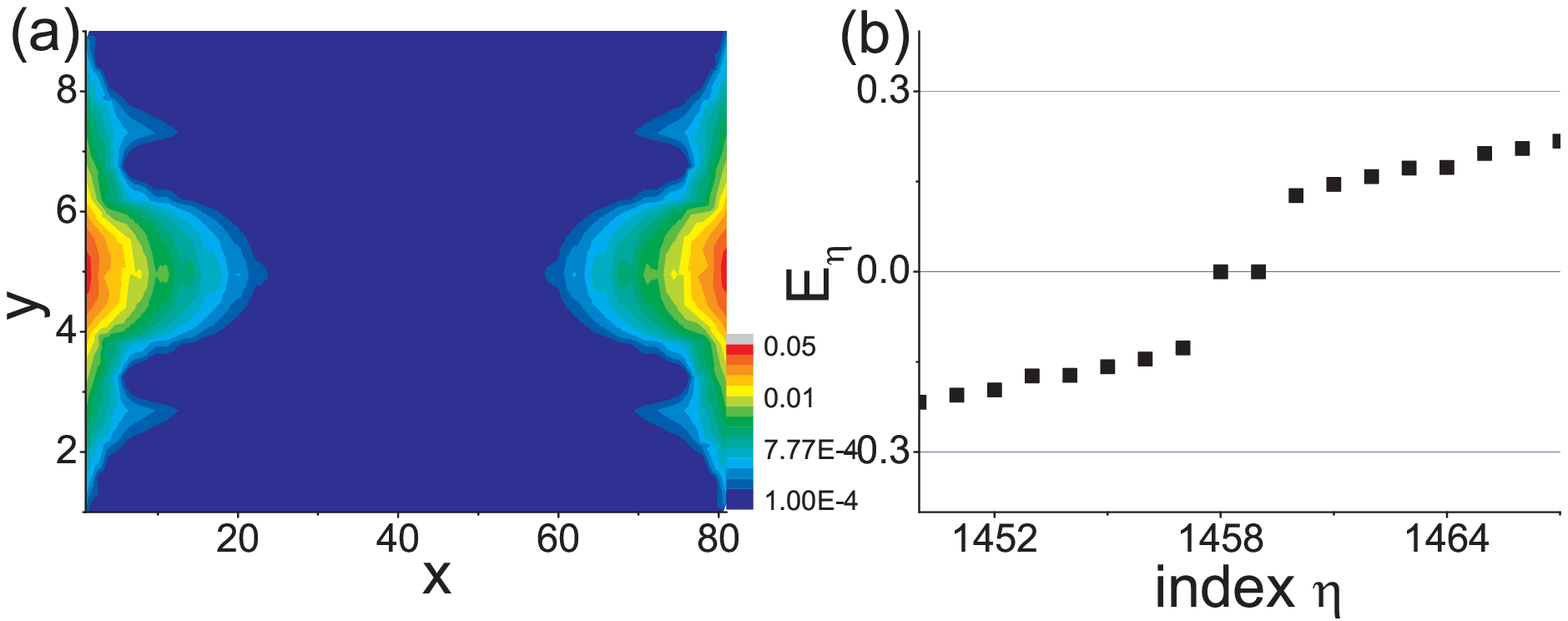} \caption{(a) Density of the zero energy mode (Log scale). (b). Quasiparticle energy spectrum. $h_{\Vert}=0.2t$,
the other parameters are the same as Fig. 2 in the main part of the
paper.}

\label{f2}
\end{figure}

In this Supplemental Material, we show the existence of MFs in a single 1D topological
chain in the 2D optical lattice with an additional in-plane Zeeman field $h_{\Vert}$. $H_{SOC}=\alpha \sum\nolimits_{\mathbf{i}}(C_{\mathbf{i},\uparrow }^{\dagger
}C_{\mathbf{i}-\hat{x},\downarrow }-C_{\mathbf{i},\uparrow }^{\dagger }C_{%
\mathbf{i}+\hat{x},\downarrow }+H.c.)+h_{z}\sum_{\mathbf{i}}(C_{\mathbf{i}%
,\uparrow }^{\dagger }C_{\mathbf{i},\uparrow }-C_{\mathbf{i},\downarrow
}^{\dagger }C_{\mathbf{i},\downarrow })
+h_{\Vert}\sum_{\mathbf{i}}(iC_{\mathbf{i},\uparrow}^{+}C_{\mathbf{i},\downarrow}+H.c.)$,
where the last term represents the in-plane Zeeman field. The self-consistent
BdG results are present in Fig. \ref{f2}.

\end{document}